# Ruling out Multi-Order Interference in Quantum Mechanics

Urbasi Sinha[1], Christophe Couteau[1,2], Thomas Jennewein[1], Raymond Laflamme[1,3] and Gregor Weihs[1,4]

[1]*Institute for Quantum Computing and Department of Physics and Astronomy, University of Waterloo, 200 University Avenue West, Waterloo ON N2L 3G1, Canada.* [2]*Laboratoire de Nanotechnologie et d'Instrumentation Optique, Université de Technologie de Troyes, 12 rue Marie Curie, 10 000 Troyes, France.* [3]*Perimeter Institute for Theoretical Physics, 31 Caroline Street North, Waterloo N2L 2Y5, Canada.* [4]*Institut für Experimentalphysik, Universität Innsbruck, Technikerstraße 25, 6020 Innsbruck, Austria.*

Corresponding authors: usinha@iqc.ca, gregor.weihs@uibk.ac.at

**Quantum mechanics and gravitation are two pillars of modern physics. Despite their success in describing the physical world around us, they seem to be incompatible theories. There are suggestions that one of these theories must be generalized to achieve unification. For example, Born's rule, one of the axioms of quantum mechanics could be violated. Born's rule predicts that quantum interference, as shown by a double slit diffraction experiment, occurs from pairs of paths. A generalized version of quantum mechanics might allow multi-path, i.e. higher order interferences thus leading to a deviation from the theory. We performed a three slit experiment with photons and bounded the magnitude of three path interference to less than $10^{-2}$ of the expected two-path interference, thus ruling out third and higher order interference and providing a bound on the accuracy of Born's rule. Our experiment is consistent with the postulate both in semi-classical and quantum regimes.**

Born's interpretation (1) of the wave function $\psi(\vec{r},t)$ for a quantum mechanical state stipulates that the probability density to find a particle at position $\vec{r}$ and at time $t$ is given by:

$$P(\mathbf{r},t) = \psi^*(\mathbf{r},t)\psi(\mathbf{r},t) = |\psi(\mathbf{r},t)|^2 \qquad (1)$$

A double slit diffraction experiment is a direct consequence of this rule and the probability to detect a particle at $\vec{r}$ after passing through an aperture with two slits *A* and *B*, is given by $P_{AB}(\mathbf{r}) = |\psi_A(\mathbf{r}) + \psi_B(\mathbf{r})|^2 = |\psi_A|^2 + |\psi_B|^2 + \psi_A^*\psi_B + \psi_B^*\psi_A = P_A + P_B + I_{AB}$, where we have omitted the position argument for brevity and defined $P_i$ to be the probability with only slit *i* (*i* = *A*, *B*) open. The corresponding (second order) interference term can be defined as

$$I_{AB} := P_{AB} - (P_A + P_B) = P_{AB} - P_A - P_B, \qquad (2)$$

Within quantum mechanics, adding more paths, i.e. slits, doesn't add higher complexity. For three slits *A*, *B*, and *C* (Fig. 1), we find $P_{ABC} = P_A + P_B + P_C + I_{AB} + I_{AC} + I_{BC}$. Therefore by Born's rule and its square exponent (Eq. 1), interference always occurs in pairs of possibilities and is defined as the deviation from the classical additivity of the probabilities of mutually exclusive events (2). These possibilities can be associated with any degree of freedom, such as spatial paths, energetic states, angular momentum states etc. Even if multiple particles are involved, interference occurs in pairs of possibilities. Consequently we define the third order interference term $I_{ABC}$ for a three-path configuration (mutually exclusive), as the deviation of $P_{ABC}$ from the sum of the individual probabilities and the second order interference terms:

$$I_{ABC} := P_{ABC} - (P_A + P_B + P_C + I_{AB} + I_{BC} + I_{AC}) = \\ = P_{ABC} - P_{AB} - P_{BC} - P_{AC} + P_A + P_B + P_C. \quad (3)$$

A physical system with such probability terms is three-path interference of a photon sent through a mask with three slits (Fig. 1). Note that the definitions in Eq. 2 and Eq. 3 are the first terms in an infinite hierarchy of interference terms (2).

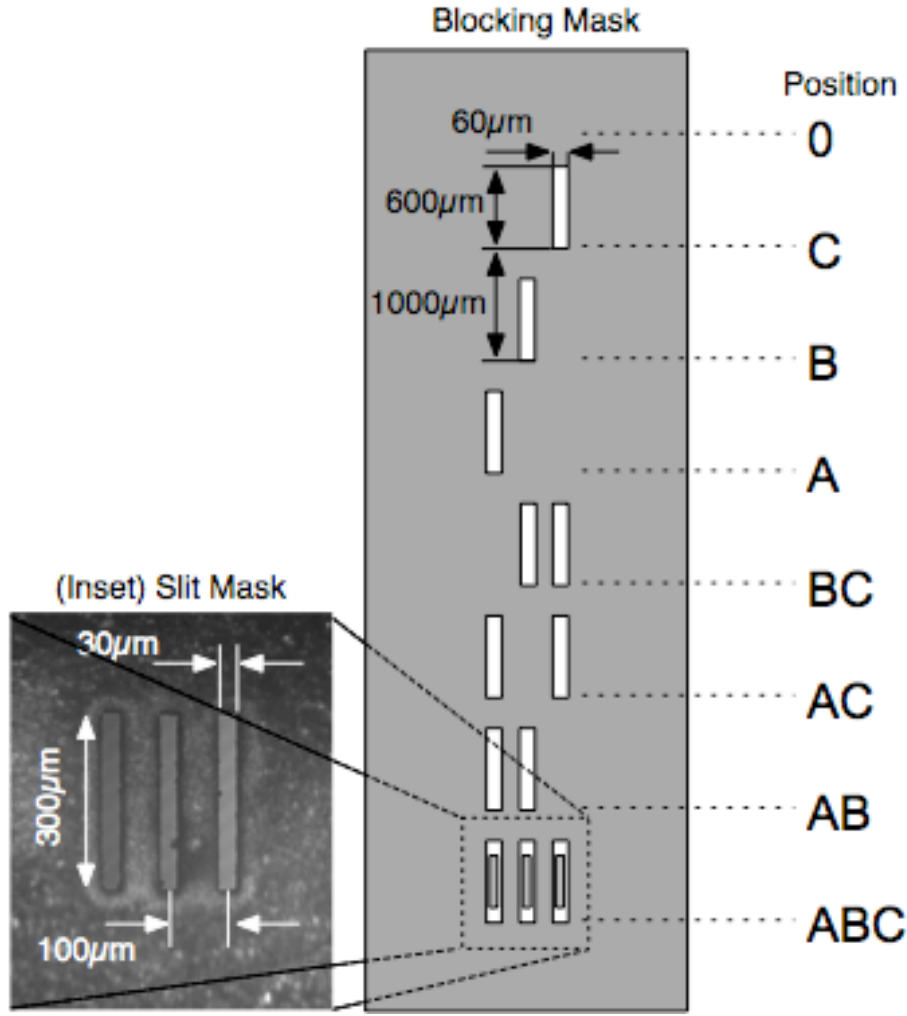

Figure 1. Arrangement and dimensions of the slits used in the experiment: The blocking mask has open apertures depending on the measured slit combination according to Eq. 4. (Inset) Image of the triple slit mask.

The non-zero interference term $I_{AB}$ is expected in all wave theories including quantum mechanics (3, 4). The next higher order, i.e. three-path interference term $I_{ABC}$, will be zero in all wave theories with a square-law relation between the field energy (or probability density) and field amplitude, which is the case in quantum mechanics with Born's rule. Moreover, if there is no interference at a certain level in the hierarchy, the higher order terms must vanish as well (2).

Our aim is to establish experimentally whether the value of $I_{ABC}$ is different from zero. We measure all seven probability terms of Eq. 3 plus the probability $P_0$ of detecting particles when all slits are closed. $P_0$ represents the probability of the empty set in an abstract definition, or a background signal in the experiment. The eight terms are obtained by sending optical photons through three slits, which can be opened or closed individually (see Fig. 1 for the slit details and Fig. 2 for the set-up). A double slit experiment could be used to test Born's rule, but then one would have to measure the non-zero double slit interference term and compare it with the theoretical prediction. This would be sensitive to experimental parameters such as slit dimensions, wavelength of incident photons and distance between detector and slits; each with its attendant error. In contrast, we expect the three path interference term $I_{ABC}$ to be zero, with the advantage of being independent of many experimental parameters, thus enabling a more precise null test for Born's rule.

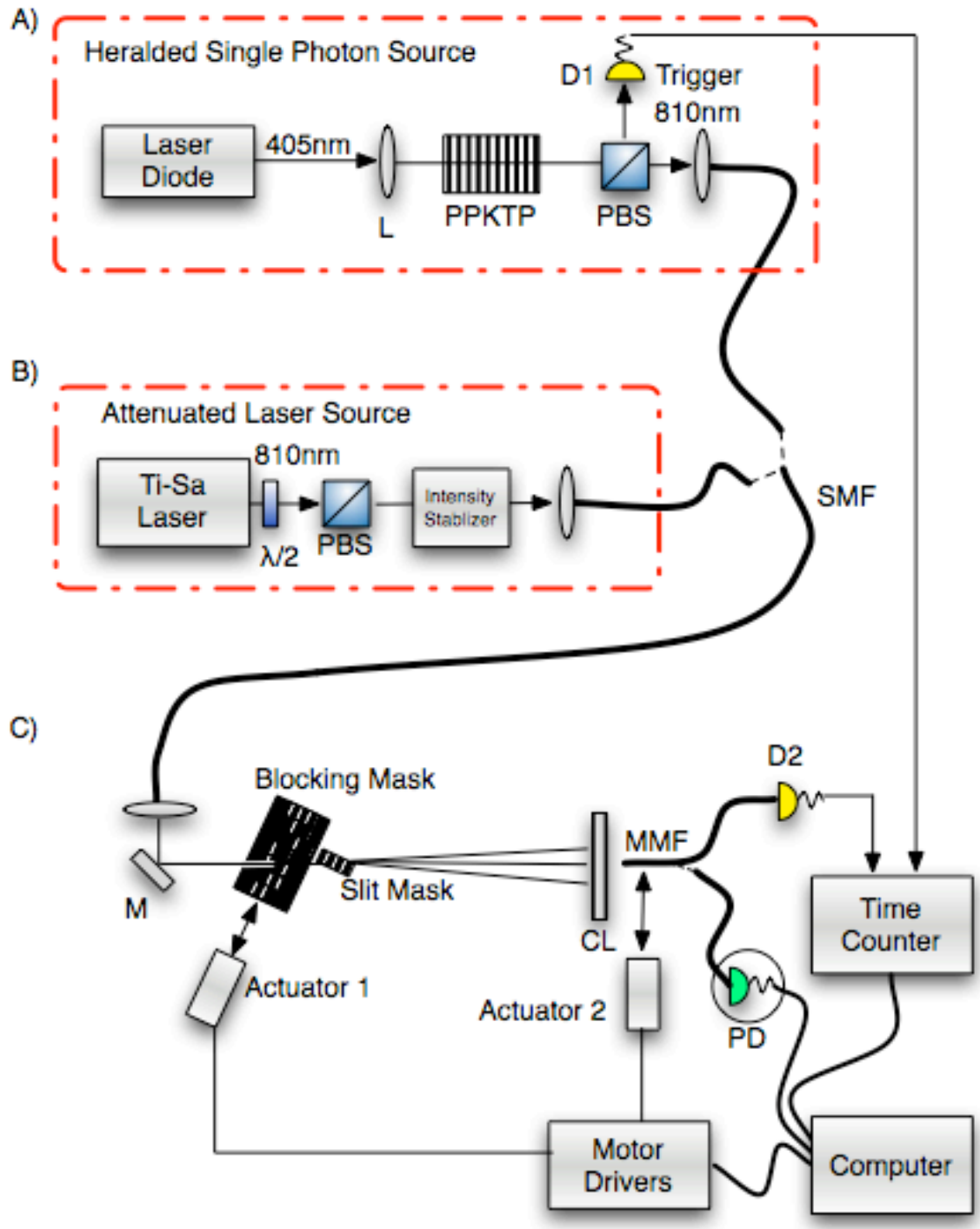

Figure 2: Experimental set-up used for the measurement of $\kappa$. A) Creation of heralded single photons from a periodically poled KTP (PPKTP) non-linear crystal pumped by a 405 nm laser diode. Parametric down-converted photons are emitted as pairs at 810 nm, and are coupled into a single mode fibre (SMF). Photon detection (D1) in the Trigger output heralds a single photon then sent through the diffraction slits. B) A pulsed titanium-sapphire (Ti-Sa) laser is attenuated and coupled into a SMF. The attenuation is realised by the combination of a half wave-plate ($\lambda/2$) and a polarising beam-splitter (PBS), combined with neutral filters and an intensity stabiliser. C) Schematic of the 3 slit experiment where the photons from the source go through the movable blocking mask with the eight combinations and then through the slit mask which has the three slits cut into it. We keep the slit mask stationary whereas the blocking mask consists of bigger and wider slits that open up the various slit combinations

as it moves up and down. This way, we ensure that the same set of slits is used for measuring the different combinations thus eliminating any dependence on the slit properties. The diffracted light is condensed vertically with a cylindrical lens (CL) onto a multimode fibre (MMF, core size 62.5μm) ~180 mm from the slits. This fibre (moveable along the diffraction pattern) acts as an aperture to probe the interferences. The collected photons are detected either with an avalanche photodiode (D2) whose signals are recorded with a time counter or with an optical power meter (PD), both connected to a computer. For heralded single photons, detections are conditioned on the detection of a trigger photon.

We measure the terms in Eq. 3 as well as $P_0$ which accounts for the inevitable detector noise and background signal. The measured quantity $\varepsilon$ based on Eq. 3 is given by

$$\varepsilon = p_{ABC} - p_{AB} - p_{AC} - p_{BC} + p_A + p_B + p_C - p_0 \tag{4}$$

Here, $p \propto P$ of Eq. 3 and refers to the measured number of photons (or optical intensity, proportional to the photon number) in the various slit combinations. To give a scale to the size of a potential deviation from Born's rule, we define a normalized variant of $\varepsilon$ called $\kappa$ (Fig. 3),

$$\kappa \equiv \frac{\varepsilon}{\delta}, \tag{5}$$

where

$$\delta = |I_{AB}| + |I_{BC}| + |I_{CA}| =$$
$$= |p_{AB} - p_A - p_B + p_0| + |p_{BC} - p_B - p_C + p_0| + |p_{CA} - p_A - p_C + p_0|. \tag{6}$$

Here, $\delta$ is the sum of the absolute values of the double slit interference terms and $\kappa$ can be seen as the ratio of an unexpected three-path interference term to the expected two-path interference term. If $\delta = 0$, then $\varepsilon = 0$ trivially and one deals with classical probabilities instead of quantum behaviour. Thus a non-zero $\delta$ ensures that we are in a quantum mechanical regime. In an experiment, we never measure probabilities directly, but only absolute frequencies of photon occurrences. The quantity $\kappa$ is independent of the total particle flux onto the slits as long as it is constant in time.

To measure $\kappa$ in various optical power regimes, we used different types of photon sources. Figure 2 shows details of the experimental set-up. We used a laser attenuated to a power level of a few microwatts down to ~200 fW (single-photon level) as well as heralded single photons (~40000 photons/s) created by spontaneous parametric down-conversion (5).

At the photon counting level, the detection mechanism is based on a silicon avalanche photodiode (APD) and thus the particle-like nature of light is incorporated in the experiments. At the microwatt level, a series of measurements was performed with a standard optical power meter, using a silicon photo-diode. The power meter measurements investigated the optical regime in which particle character is not of concern. In all cases we performed a large number of measurements at fixed points in the diffraction pattern (Fig. S1 in (5)). In addition, we have also performed measurements to check the variation of $\kappa$ as a function of detector position. Born's rule would predict that $\kappa$ should be independent of detector position. However, systematic errors may vary with the position and therefore are seen to bring a variation in the measured value of $\kappa$ at different detector positions even in our experiment. Nevertheless, the mean $\kappa$ is within the bounds set by the attendant errors at each such detector position.

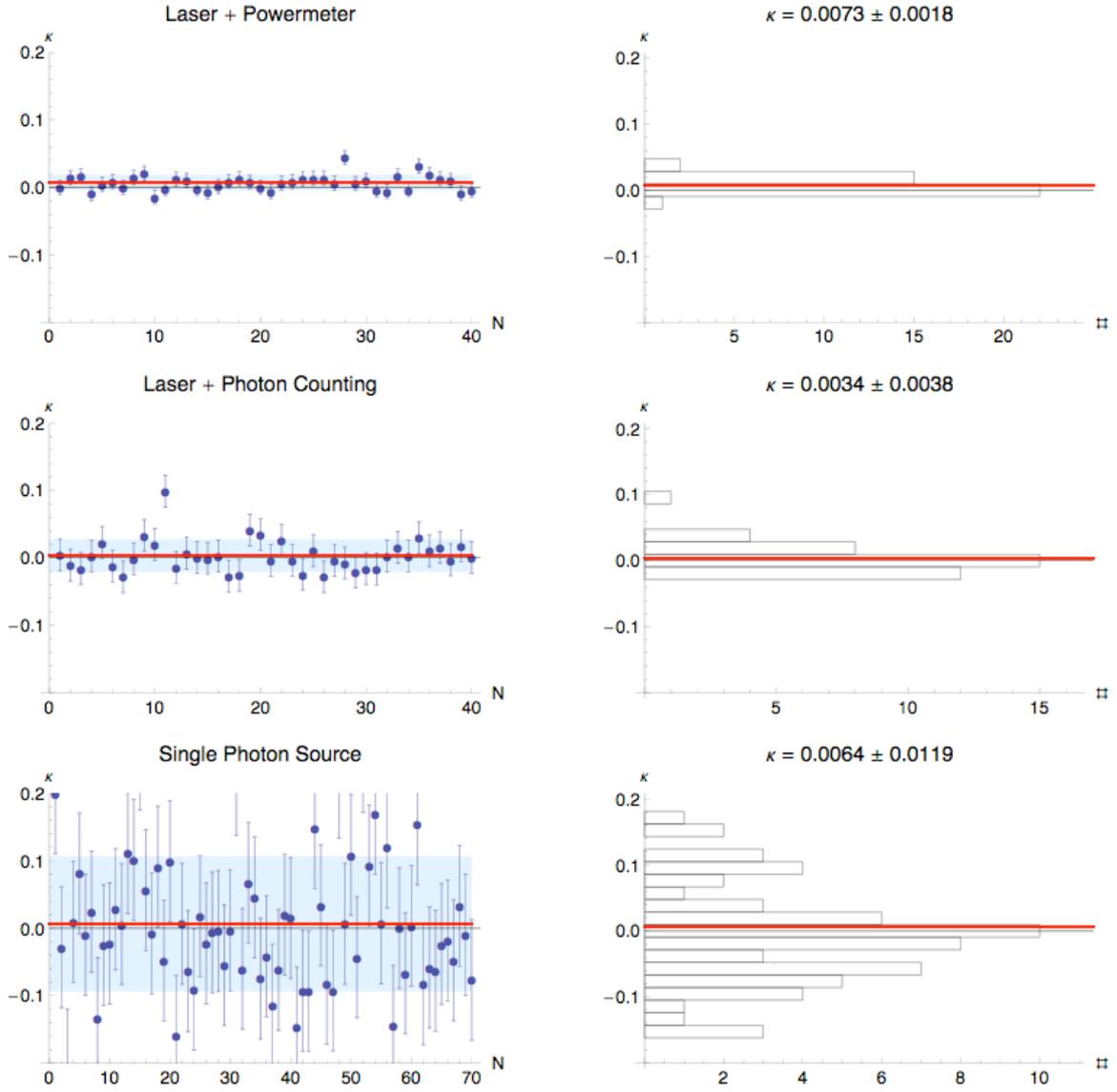

Figure 3: Typical distributions of $\kappa$ measured at the central maximum of the triple slit interference pattern. Top row shows data using a laser source and a power meter for detection with $\kappa = 0.0073 \pm 0.0018$. Middle row represents data with the laser but attenuated to single photon level and an APD for detection resulting in $\kappa = 0.0034 \pm 0.0038$. Bottom row data was taken using the heralded single photon source (HSPS) and an APD for detection resulting in $\kappa = 0.0064 \pm 0.0119$. The Allen variance of $\kappa$ was used to plot the error bars. The horizontal red lines represent the mean $\kappa$ values and the blue shaded regions represent a band of one standard deviation of the distribution of $\kappa$ values around the mean.

The typical distributions of measured values of $\kappa$ are shown in Fig. 3, with photon streams from a laser attenuated to different levels (top and middle rows) and from a heralded single photon source (bottom row). $\kappa$ is calculated from the measured interference intensities for the eight independent slit combinations at a fixed position. The order of the eight slit combinations was chosen randomly for reducing systematic influences on $\kappa$ caused by slow variations of the photon flux. Each combination in a run was measured for a certain photon-count integration time and up to 100 runs were cycled in order to obtain a statistically significant sample of $\kappa$ values. Among the many positions in the diffraction pattern, we chose the central maximum of the triple slit combination (yielding the maximum number of coincidence photon counts) to obtain our data (5). For the single photon source, we measured at each slit combination until the trigger count reached 30 million, which was a good compromise between accumulating a statistically significant number of coincidences for the different slit combinations and ensuring a low drift of the photon source between measurements.

With a null experiment, a very careful analysis of random and systematic errors must be undertaken, as our bound on the amount of three-path interference will be directly related to the level of experimental uncertainty. Among the random errors in our setup, thermal and acoustic fluctuations cause the source fluxes to vary in time. In addition, detection efficiency and optical alignment can change. In particular there will be some mechanical vibration of the thin (25 µm) slotted steel membrane apertures, causing a variable slit transmission due to near field diffraction. In addition, for power meter measurements the instrumental error is added to the above error sources, whereas for photon counting, the Poissonian distributed counting error is the dominating fluctuation. Because of the random nature of the individual errors we used Gaussian error propagation to estimate the error of $\kappa$, where we use the standard variances of the individual measurement values calculated from a large number of repetitions of the experiments. In some cases where we observe a drift in the rates, we found the Allen variance of the values to be a better estimator for error propagation. This is justifiable because $\kappa$ is calculated from eight measurements taken in direct succession and the variance between subsequent samples of each quantity $p_A$, $p_B$, etc. is therefore the most suitable error estimator.

Once we understand the random errors we can characterize the systematic errors. Our experiment and the measurement of $\kappa$ are convenient, as they neither require the slits to be identical, nor require the transmission values to be perfectly 1 and 0. On the other hand what matters is the absence of correlation or systematic variation in how the slits behave while switching between slit combinations. Note that the size of the slits and the wavelength make independent shutters difficult to insert and we used a static-opening mask plate in front of the actual slits for blocking and unblocking the individual slits. Our approach can potentially introduce unwanted correlations between the switching of different combinations. This occurred in our case, a fault in the blocking mask in the *BC* combination caused opening *B* to be shifted off its nominal position by 8 µm. At zero distance between opening and slit masks this shift would not affect the transmittance of the diffracting slit. However, with the finite separation of 50 µm between the two masks, it does matter. Using a two-dimensional finite-difference-time-domain simulation of the light field between the wider opening and narrower slit, we found that this lateral misalignment could change the effective slit transmission by ± 3%, depending on the relative position. Using this to adjust the slit transmittance value for slit *B* in the combination *BC* leads to a value of $|\kappa| = 0.01$. Bringing the masks closer would reduce this error, but the thin membranes tend to stick together when they come closer than 50 µm.

Another major systematic error is detector nonlinearity. There is no perfect detector and efficiency and nonlinearity will always be finite. For our commercial power-meter, the nonlinearity is 0.5% for all ranges. For the worst case, at the central maximum of the diffraction pattern, this results in a systematic error of $|\kappa| = 0.003$. For photon counting detectors, there is always a finite dead-time during which they are blind to photons. Depending on the flux, this results in a saturation effect, that is, the detector response does not follow the square law, but has a deviation potentially indistinguishable from a violation of Born's rule. Therefore, to keep non-linearity errors negligible, we used count rates below 100,000/s.

Combining the various error sources, our particular setup enables us to bound the measurements of $\kappa$ to better than 0.01, thus providing a bound on the accuracy of Born's rule. This is in good agreement with the values measured for single photons $\kappa = 0.0064 \pm 0.0120$ and the attenuated laser beams $\kappa = 0.0073 \pm 0.0018$ (power meter measurement) and $\kappa = 0.0034 \pm 0.0038$ (APD measurement) from Fig. 3.

Note that any significant non-zero observation of $I_{ABC}$ would imply that Born's rule doesn't strictly hold. The consequences of detecting even a small amount of three-way interference (by deviating from the quantum mechanical null prediction) would be tremendous. A modification to Born's rule that leads to multi-order interference would have repercussions on the allowable dynamics. In particular, if probability must be conserved then Schrodinger's equation would likely have to be modified as well. Nonlinear extensions to quantum mechanics are one way to generalize it (6-8) and there have been efforts to test these nonlinearities (9-12). However, in such experiments, a model was assumed for the nonlinear variant of the Schrödinger equation and efforts concentrated on estimating the coefficient of the nonlinearity. In contrast, we present a dedicated test for Born's rule and we are confirming it within our experimental limitations without depending on specific nonlinear extensions of quantum mechanics.

We are able to bound the magnitude of the third order interference term to less than $10^{-2}$ of the regular expected second order interference, at several detector positions. Thus, our experiment is able to rule out the existence of third-order interference terms (and, in effect, any higher order interference terms) up to this bound. This bound on the accuracy of Born's rule is relevant for theoretical attempts to derive it (13) as well as for the generalization of quantum mechanics. Among the consequences of such a generalized theory, it would require more detailed specifications of the quantum system (14) and could modify computational complexity by allowing one to distinguish between orthogonal states thus breaking quantum cryptography and making quantum computing more powerful i.e. super-quantum computing (15). The triple slit experiment is a simple and very natural system to investigate three-path interference, but it is not the only possible implementation; any configuration of three mutually exclusive quantum paths can be used in such a test. Although our implementation did not lead to observation of any deviations, such future tests with other systems might indeed lead to improved accuracies on the bound of $\kappa$. It would be interesting to perform tests with other types of particles such as neutrons (16, 17), $C_{60}$ molecule interference (18), electron interference with slits or potential wells (19, 20) or using a system with a global wave function such as a Bose-Einstein condensate (21, 22).

(23) U.S., C.C., and T.J. performed the experiments, C.C., R.L., and G.W. conceived of and supervised the experiment and developed the theoretical aspects. G.W. and U.S. performed the modelling, data and error analysis. All authors contributed to the writing of the paper. We would like to thank H. Hübel, R. Kaltenbäck, A. J. Leggett, A. Sinha, R. D. Sorkin and A. Zeilinger for valuable discussions and Z. Medendorp and I. Söllner for technical assistance earlier on in the experiments. We acknowledge financial support from Natural Science and Engineering Research Council, Canada Foundation for Innovation, European Research Area, QuantumWorks, Ontario Centres of Excellence and Canadian Institute for Advanced Research.